\documentclass{article}
\usepackage[utf8]{inputenc}
\usepackage{authblk}
\usepackage{setspace}
\usepackage[margin=1.25in]{geometry}
\usepackage{graphicx}
\usepackage{subcaption}
\usepackage{amsmath}
\usepackage{lineno}
\usepackage{latexsym}
\usepackage{booktabs}
\usepackage{color}
\usepackage{soul}

\RequirePackage[colorlinks=true, allcolors=blue]{hyperref}

\usepackage[style=ieee, 
citestyle=numeric-comp,
sorting=none]{biblatex}
\title{Compact  InGaAs/InP single-photon detector module with ultra-narrowband interference circuits }

\begin{document}
\author[1†]{Zhengyu~Yan}
\author[1†]{Tingting~Shi}
\author[1*]{Yuanbin~Fan}
\author[1]{Lai~Zhou}
\author[1]{Zhiliang~Yuan}

\affil[1]{Beijing Academy of Quantum Information Sciences, Beijing 100193, China}
\affil[$\dag$]{These authors contributed equally to this work.}
\affil[*]{Corresponding author. Email: fanyb@baqis.ac.cn}

\date{}

\onehalfspacing

\maketitle

\begin{abstract}
Gated InGaAs/InP avalanche photodiodes are the most practical device for detection of telecom single photons arriving at regular intervals.  
Here, we report the development of a compact single-photon detector (SPD) module measured just $8.8 \times 6 \times 2$~cm$^3$ in size and fully integrated with driving signal generation, faint avalanche readout and discrimination circuits as well as temperature regulation and compensation.  The readout circuit employs our previously reported ultra-narrowband interference circuits (UNICs) to eliminate the capacitive response to the gating signal. 
We characterize a UNIC-SPD module with a 1.25 GHz clock input and find its performance comparable to its counterpart built upon discrete functional blocks.  Setting its detection efficiency to 30~\% for 1550~nm photons, we obtain an afterpulsing probability of 2.4~\% and a dark count probability of $8\times 10^{-7}$ per gate under 3 ns hold-off time.  We believe that UNIC-SPDs will be useful in important applications such as quantum key distribution.

\end{abstract}

\section{Introduction}

A single-photon detector (SPD) is sensitive to incidence of individual quanta of light and has many applications in  photonics, such as fluorescence measurements~\cite{Damalakiene2016,Albertinale2021}, laser ranging~\cite{Wehr1999, McCarthy2013,Zheng2021}, optical time-domain reflectometer~\cite{Eraerds,Li}, and quantum optics experiments \cite{Korneev2007}. Near-infrared SPDs at the telecommunication wavelength of 1550~nm are indispensible for fiber-optic quantum key distribution (QKD)~\cite{Liang2014,Comandar2014,yuan2018}, 
with choices~\cite{Eisaman2011} including cryogenic superconducting nanowire single-photon detectors and electrically-cooled InGaAs avalanche photodiodes (APDs).  Between them, InGaAs have practical advantages for compactness, cost and ease in operation.  

Under Geiger mode, avalanche current flow in an APD is self-sustaining and must therefore be quenched so as to rearm the device for detection of subsequent photons.  Traditional passive quenching method, which works well for Si APDs \cite{Kim2011}, has to be used in conjunction with an excessively long dead time on the order of 10~$\mu$s for InGaAs APDs in order to suppress the afterpulsing to an acceptable level \cite{korzh2014}, thus placing a severe limit on the maximum count rate.  Fortunately, this limit does not apply to gigahertz-clocked gated Geiger mode~\cite{Namekata2006,Yuan2007}
which allows a count rate of up to 1~GHz~\cite{patel2012}. 
As a drawback, APD's strong capacitive response to sub-nanosecond gating has to be rejected through 
purpose-designed
readout circuits~\cite{Namekata2006,Yuan2007,Restelli2013,He2017,Fan2023,HeDeyong2023,} so as to enable detection of weak photon-induced avalanches.  
Rapid gating 
adds challenges to modularization and miniaturisation, 
but which is a necessary step to serve a wide range of applications. Nevertheless, successful efforts have been made to integrate InGaAs APDs into state-of-the-art QKD systems~\cite{yuan2018,dynes2019}, while the compactest module was achieved with a monolithically integrated readout circuit~\cite{Jiang:17,Jiang2018}.

\begin{table}[t]
    \centering
    \caption{\textbf{Comparisons of the UNIC-SPD module with our previous detector~\cite{Fan2023} built with bulk driving and discrimination electronics, and with prior state-of-the-arts of InGaAs SPD modules~\cite{Jiang:17,Jiang2018} gated at 1.25~GHz clock.} }
    
    \resizebox*{\textwidth}{!}{
        \begin{tabular}{cccccccc}
       \toprule
          & Dimension~(cm$^3$) &$\eta _{net}$~(\%) & $P_A$~(\%) & $P_D$~(gate$^{-1}$)  & T~(℃)&  Hold-off time~(ns) & Count rate (MHz) \\
        \midrule
            Previous UNIC detector~\cite{Fan2023} & Not applicable & 30  & 2.0  & $5.1\times 10^{-6}$  & -15 &   - & 700  \\
            \hline   
           This work & 8.8 × 6 × 2 & 30 & 2.4 & $8.0\times 10^{-7}$  & -15 &  3 & 220  \\
            Jiang \textit{et al.}, 2017~\cite{Jiang:17} & 12 × 7 × 5 &27.5 & 9.1 & $9.6\times 10^{-7}$ & -50  &  100  &   - \\
            Jiang \textit{et al.}, 2018~\cite{Jiang2018} & 13 × 8 × 4  &30 & 8.8 & $3.5\times 10^{-7}$  & -30 &  100  &   -  \\
        \bottomrule
        \end{tabular}    
    }
    \label{tab:Table 1}
\end{table}

We have recently developed a novel readout circuit~\cite{Fan2023} that incorporates a surface acoustic wave (SAW) filter into an asymmetric  radio-frequency (RF) Mach-Zehnder interferometer, referred to as ultra-narrowband interference circuit (UNIC), and realized exceptional performance for narrow-band rejection of the APD  capactive response.  Thanks to the long group delay of the SAW filter,  the UNIC interferometer can 
produce an ultra-narrrow band rejection with a manufacturing tolerance easily achievable in the RF track lengths. The UNIC can provide a wide and continuous pass band in the frequency domain and therefore brings little distortion into the avalanche signal.  
Here, we report our development of a standalone InGaAs SPD module that fully integrates driving and readout electronics as well as temperature regulation and compensation.  As shown in Table~\ref{tab:Table 1}, its dimension is measured just 8.8 × 6 × 2 $\rm {cm}^3$ and is nearly a factor of 4 smaller in volume than the existing compactest detector module that uses a monolithically integrated readout circuit~\cite{Jiang2018}.
Important for practical applications, we implement a field-programmable-gate-array (FPGA) control to compensate for temperature drift  and thus ensure stable operation of the UNIC-SPD module over varying ambient conditions. 

\section{Design and Fabrication}

\begin{figure}[h]
    \centering       \includegraphics[width=0.9\textwidth]{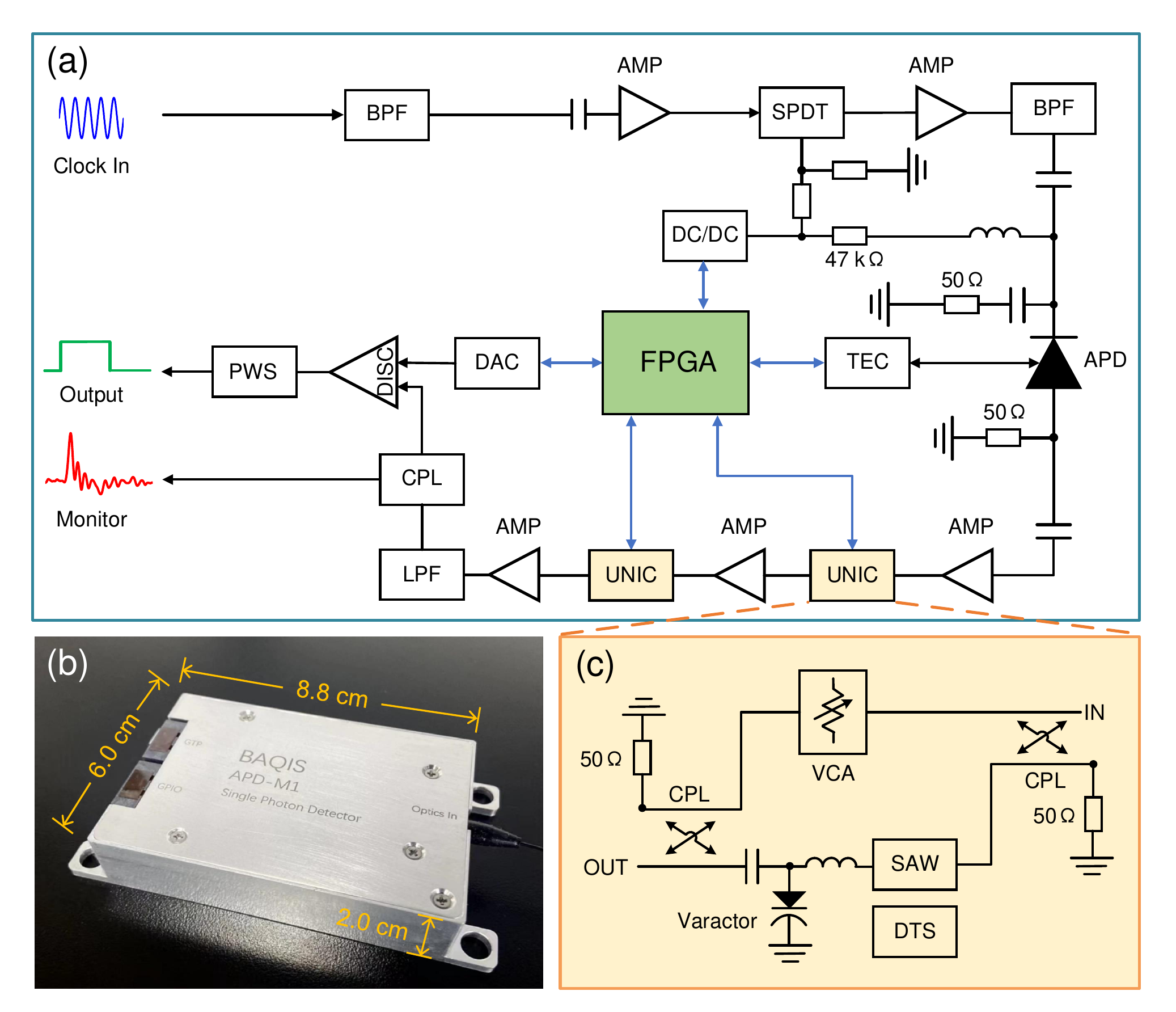}
    \caption{(a) Schematic setup of the fully integrated InGaAs UNIC-SPD module; (b) Photo of a UNIC-SPD module; (c) Schematic for the UNIC interferometer.
    BPF: band pass filter; AMP: amplifier; SPDT: single pole double throw switch;  DC/DC: DC/DC converter; APD: avalanche photodiode;  TEC: thermo electric cooler; FPGA: field programmable gate array; DAC: digital-to-analog converter; DISC: discriminator; PWS: pulse width shaper; CPL: coupler; LPF: low pass filter;  UNIC: ultra-narrowband interference circuit;  VCA: voltage controlled attenuator; SAW: surface acoustic wave; DTS: digital temperature sensor.
    }
    \label{Fig.1}
\end{figure}

Figure~\ref{Fig.1}(a) shows the schematic of internal functional blocks of our fully integrated InGaAs/InP UNIC-SPD module.  The module has a dimension of $8.8 \times 6.0 \times 2.0$~cm$^3$, as shown in  the Fig.~\ref{Fig.1}(b).  It contains a commercial fiber-pigtailed InGaAs/InP APD (Wooriro mini-flat), and its driving, readout and temperature regulation and compensation electronics.  Through a bias-Tee made of an inductor and a capacitor, the APD receives its DC bias of 71.6~V from a DC/DC converter that is powered by a 3.3~V voltage line and its 1.25~GHz sinusoidal-wave gating signal with $V_{pp}$ amplitude of up 12~V via amplifying the an external clock input.  In the DC bias path, a 47~k$\Omega$ series resistor is used in order to limit the maximum current through the APD. A single pole double throw (SPDT) switch is placed in the RF amplifiers' chain  and disables the gating signal for protection the APD when the output voltage of the DC/DC converter is below 40~V. 

The output from the APD anode consists of both probabilistic photon-induced faint avalanches and always present gate-induced capacitive response. These signals are sampled through a 50~Ohm resistor and then amplified by a first stage broadband low noise amplifier (LNA).  Amplification at this stage is desirable because it can prevent the faint avalanche signals from being attenuated to below thermal noise. 
The gate-induced capacitive response has an amplitude of approximately 420~mV$_{\text{pp}}$ (-3.5~dBm), the first stage LNA must have a sufficiently high output 1~dB compression point ($OP_{1\text{dB}}$) to ensure that its output signal will not saturate. Hence, we choose Qorvo TQP3M9038, which has a gain of 14.7~dB, an $OP_{1\text{dB}}$ of 21.6~dBm and a noise figure (NF) of 2~dB, for the first stage amplification. 
In the circuit, the output of the TQP3M9038 is 11.2 dBm lower than its $OP_{1\text{dB}}$, thereby ensuring its operation in linear amplification regime. The amplified output is further processed by a cascade of two UNICs and two further amplifiers to remove the gate-induced capacitive response and boost the remaining avalanche signal to have an amplitude of 1~V, respectively. The amplified avalanche signal is then split by a RF coupler (CPL) into two paths after passing through a low pass filter  (LPF).  One path is used for external monitoring, while the other is fed into a discriminator (DISC), followed by a pulse width shaper (PWS), to output a digital pulse (positive emitter coupled logic, PECL) of  a  fixed width of 1.5~ns .

To effectively remove the capacitive response spectrally concentrated at the gating frequency, which will mask the photon-induced avalanche signals, we employ UNIC technique and integrate two UNIC blocks in cascade on the same printed circuit board (PCB).  Over the previous version~\cite{Fan2023}, we have made a considerable improvement to allow UNICs to compensate for its ambient temperature change.  As shown in Fig.~\ref{Fig.1}(c),  we add into each UNIC a digital temperature sensor (DTS) in close proximity of the SAW bandpass filter, a varactor diode and a voltage controlled attenuator (VCA).    VCA allows fine balancing between the two interfering RF arms, while the varactor diode is used to compensate the temperature dependent group delay of the SAW device.  The attenuation and differential delay are set to enable destructive interference for the 1.25 GHz frequency component
at the UNIC output port. The UNIC actively adjusts and is able to stabilize the rejection ratio above 70~dB throughout our experimental test over a PCB temperature range of 20 $-$ 50 ℃. 

We have integrated a field-programmable-gate-array (FPGA) processor (Xilinx Artex 7) to enable real-time parameter monitoring and control.  It includes a digital temperature control loop through sensing the APD thermistor and setting the driving current to its built-in thermal-electric-cooler (TEC) and allows a temperature stability of $\pm$0.03~$^\circ$C over a setting range of -30 $-$ 30~$^\circ$C. The FPGA also reads the PCB temperature sensed by the DTS on each UNIC and adjusts the control voltage to the varactor diode to achieve real-time compensation for the varying group delay of the SAW filter.  The attenuation change by the varactor diode is balanced by adjusting the VCA on the other interferometer arm. Optimal   control voltages for the varactor diodes and VCAs for each temperature are pre-characterized and stored in a look-up table inside the FPGA.

Each UNIC-SPD module accepts a wide input voltage randge of 8 $-$ 16~V, and  its total power dissipation is less than 8~W.

\section{Module performance}
\subsection{Characterization setup}
We use a setup schematically shown in Fig.~2 to characterise the performance of a fully integrated UNIC-SPD module. A mode-locked laser emitting at 1550~nm is utilized as the light source, and it features a pulse width of 5 $-$ 10 ps and a repetition rate of 10~MHz. 
The laser output is split by a 50:50 fiber beam splitter, while one half is used for real-time monitoring of the laser power by an optical power meter (EXFO FTB-1750, ±5 \% uncertainty) while the other half is attenuated by a variable optical attenuator (VOA, EXFO FTB-3500) to an intensity level of 0.1~photons per pulse at the fiber entrance to the UNIC-SPD module. The laser provides a 10~MHz reference to a signal generator which synthesize a 1.25~GHz sinusoidal clock to the UNIC-SPD module. The UNIC-SPD detection output is recorded by a time-digital-converter (TDC, Swabian Time Tagger Ultra) and a photon counter (Stanford Research SR400) for systematic measurements of the detector dark count rate, detection efficiency and afterpulsing probability.  While the digital output is used for detector performance characterisation, an
oscilloscope is to monitor the undiscriminated analogue output of the UNIC-SPD module.   A  computer collects and processes all the measurement data, as well as setting the APD operational parameters including the APD temperature, its DC bias and the avalanche discrimination level. 
Throughout the test, we fix the APD temperature at -15~$^\circ$C, which gives a good compromise between the detector dark count and afterpulsing probabilities.  These noises have opposite dependencies on temperature.  A deeper cooling helps reduce the dark count probability, but afterpulsing will be more likely to occur. 

\begin{figure}[t]
    \centering  \includegraphics[width=.9\textwidth]{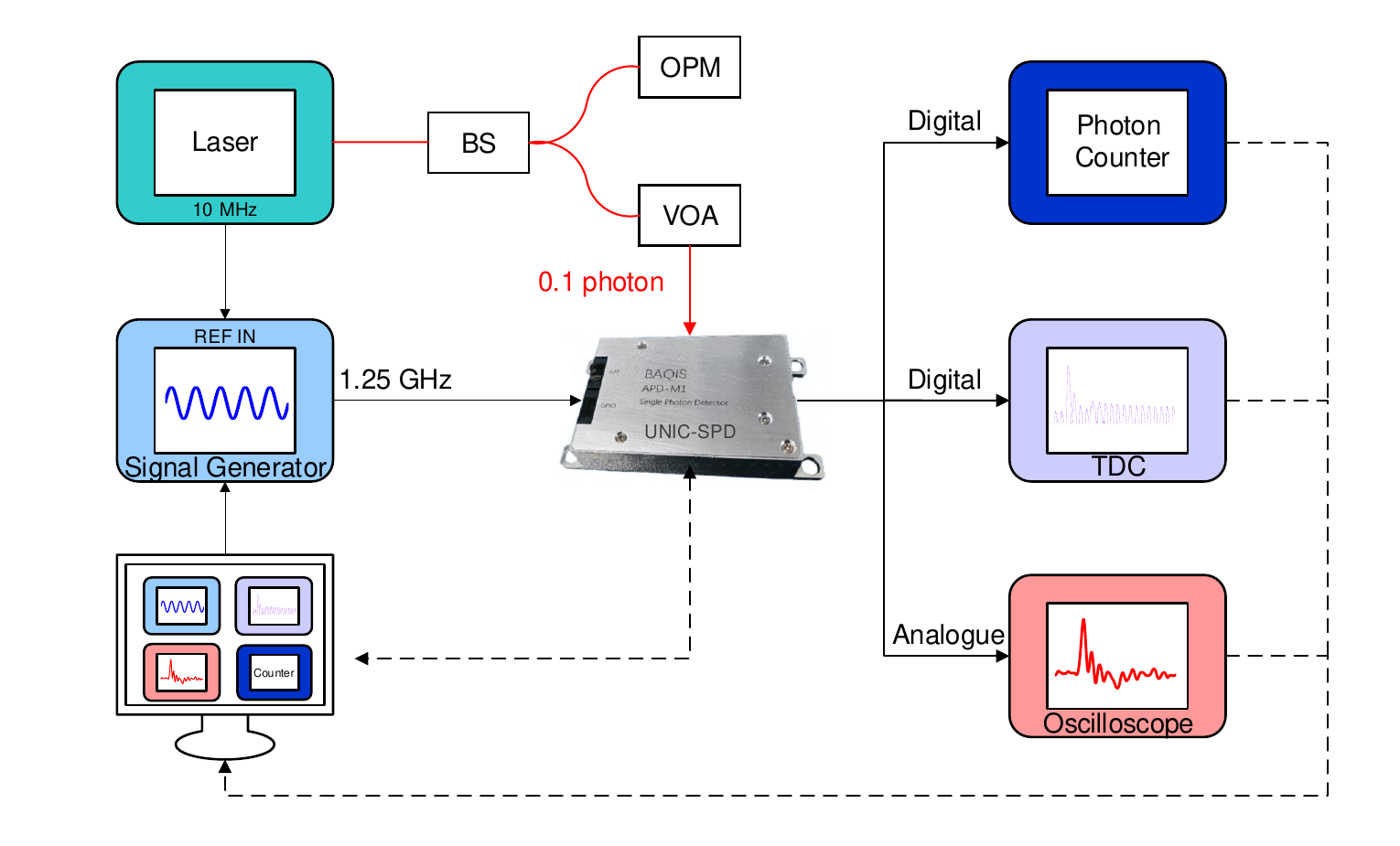}
    \caption{Characterization setup for 1.25~GHz sinusoidally gated UNIC-SPD modules. BS: beam splitter; OPM: optical power meter; VOA: variable optical attenuator; TDC: time-digital-converter.}
    \label{Fig.2}
\end{figure}

\subsection{Results and discussion}
\begin{figure}
    \centering  \includegraphics[width=0.9\textwidth]{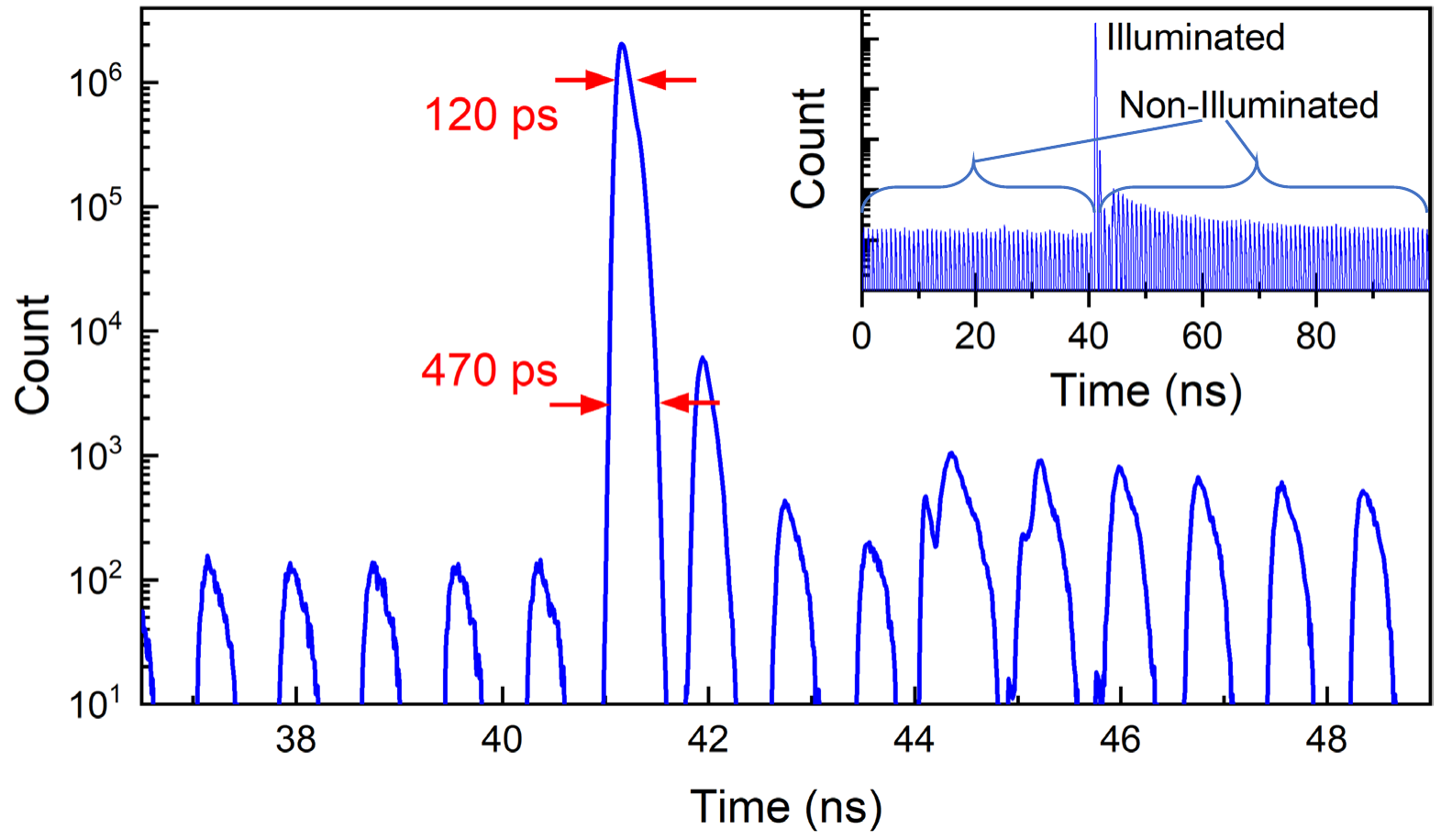}
    \caption{A histogram of photon detection events under $\eta_{net}$ = 25.8 \%. }
    \label{Fig.3}
\end{figure}

The net photon detection efficiency ($\eta _{net}$), the afterpulse probability ($P_A$) and the dark count probability ($P_D$) can be precisely measured using time-resolved photon counting. Figure~\ref{Fig.3} shows a histogram of the photon detection events under 10 MHz pulsed excitation at 0.1~photon/pulse. The main peak induced by pulse laser illuminating has a full width at half maximum (FWHM) of 120~ps, and the width of 1/1000 (30 dB width) is just 470~ps. The FWHM  deteriorates slightly as compared with previously achieved with an off-the-shelf discriminator~\cite{Fan2023}, but the whole photon detection distribution falls strictly within the gating period of 800 ps, which is vital for high speed quantum key distribution. 
The counts at non-illuminated gates are attributed to detector dark and afterpulsing events, and the corresponding count rate is 3 orders of magnitude lower than that of the illuminated gate. Non-illuminated gates between 42--44~ns exhibit suppressed counts because of the ultra-short hold-off time of 3~ns applied by the UNIC-SPD module.

From the detection events, we extract the respective count probabilities for illuminated ($P_I$) and non-illuminated gates ($P_{NI}$). 
Dark count probability ($P_D$) was measured separately by turning the laser off. The afterpulsing probability is calculated by the standard method as follow:
\begin{equation}
P_A=\frac{\left( P_{NI}-P_D \right) \cdot R}{P_I-P_{NI}},
\end{equation}
where $R = 125$ is the ratio of the detector gating frequency (1.25~GHz) to the laser illumination frequency (10 MHz). Excluding dark and afterpulse counts, the net photon detection efficiency is given by
\begin{equation}
    \eta _{net}=\frac{1}{\mu}\ln \frac{1-P_{NI}}{1-P_I},
\end{equation}
where $\mu$ is the average incident photon number per illumination pulse.

\begin{figure}[h]
    \centering
    \includegraphics[width=0.9\textwidth]{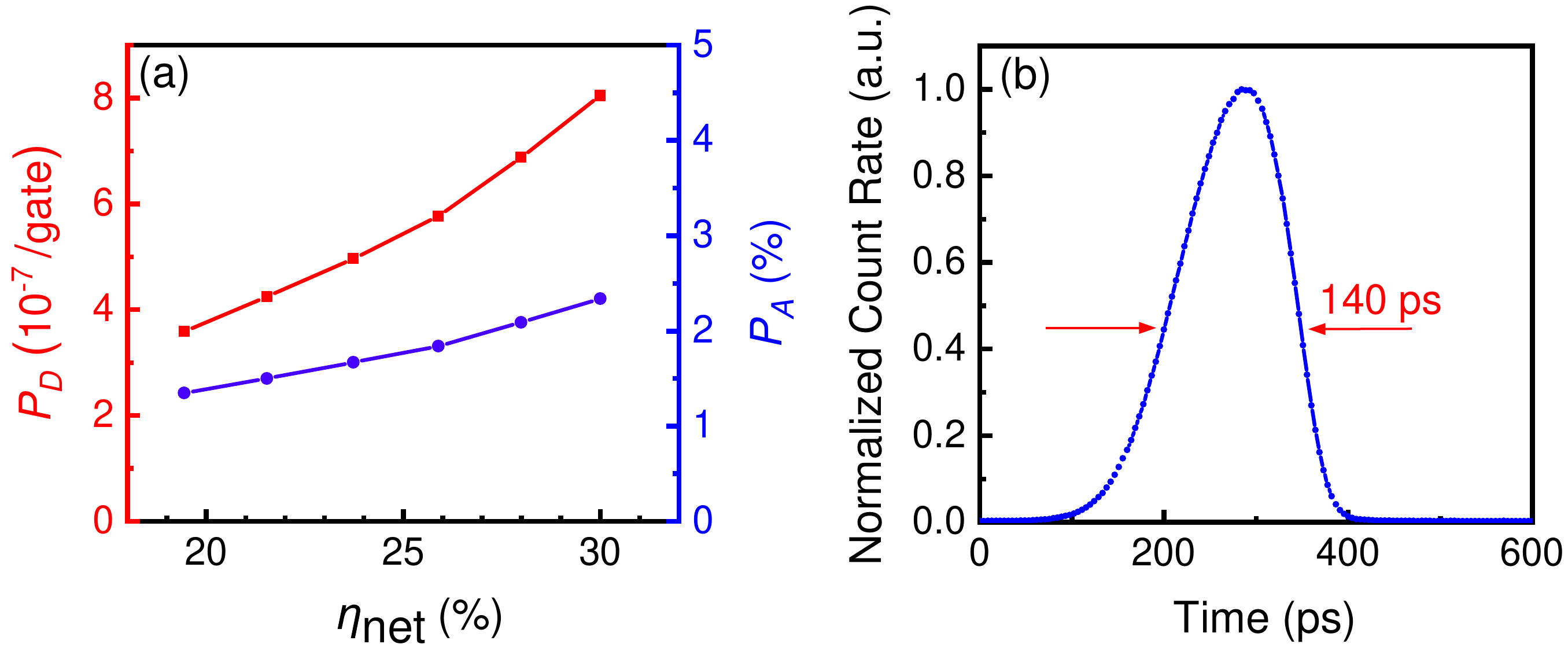}
    \caption{Performance of the UNIC-SPD module with its APD temperature set to -15$^\circ$. (a) Dark count probability ($P_D$)  and the afterpulsing probability ($P_A$) versus net photon detection efficiency ($\eta_{net}$); (b) Measured effective gating width at $\eta _{net}$ = 27.0 \%.}
    \label{Fig.4}
\end{figure}

By adjusting the amplitude of the 1.25~GHz clock input, we are able to vary the APD excess voltage and hence the detection efficiency. Figure~\ref{Fig.4}(a) plots the dark count and afterpulse probabilities as a function of net photon detection efficiency. While both dark count and afterpulse probabilities deteriorate with the photon detection efficiency, our UNIC-SPD module exhibits excellent performance. For instance, at $\eta _{net}=30\ \%$, the afterpulse probability remains below 2.4 \% and the dark count probability less than $8\times 10^{-7}$ per gate.  

Effective gate width is another important parameter for gated operation. A narrow gate helps rejection of background noise photons~\cite{patel2012coexistence}, 
but too narrow a gate is undesirable as it will require demanding synchronization. 
To determine the effective gate width, we scan the input clock delay and measure the count rate. As shown in Fig.~\ref{Fig.4}(b), the effective gating width is 140 ps FWHM at $\eta _{net}$ = 27.0 \%, which is comparable with those GHz gated APDs that are equipped with different readout circuits and have been proven usable in high-speed quantum applications~\cite{patel2012,patel2012coexistence,koehler2018best,Comandar2014}.

\begin{figure}[h]
    \centering
\includegraphics[width=0.75\textwidth]{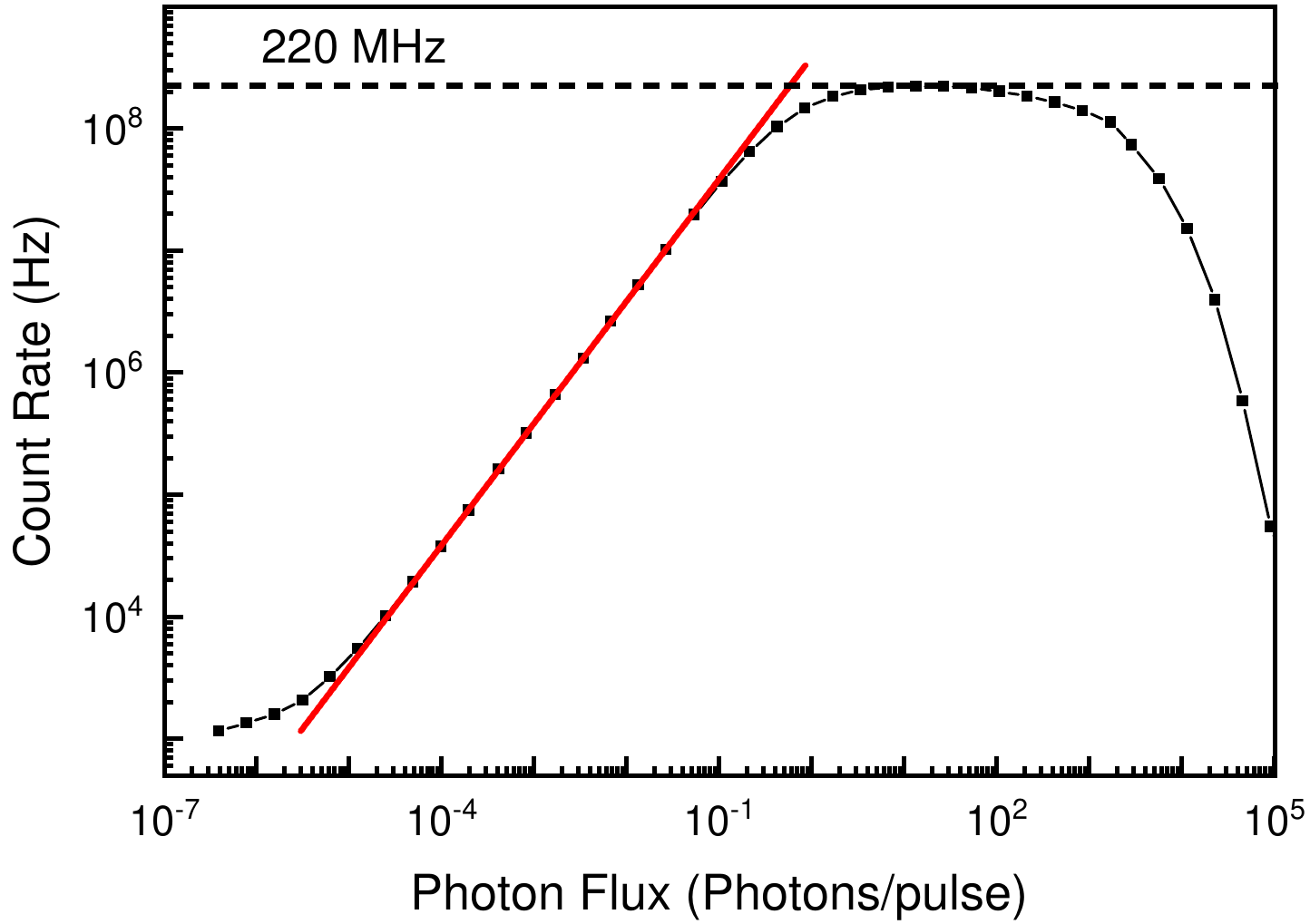}
    \caption{Count rate versus incident photon flux.}
    \label{Fig.6}
\end{figure}

Figure~\ref{Fig.6} shows the detector linearity and maximum count for the UNIC-SPD module. For illumination, we employ a distributed-feedback semiconductor laser diode, gain-switched to produce 1.25~GHz pulses of 30~ps duration.
The measured count rate starts from the dark count rate level at low fluxes and then grows linearly with the incident photon flux before saturation at fluxes greater than 0.1~photons per pulse.
The linear region spans over more than three orders of magnitude between incident fluxes of $2\times 10^{-5}$ and $10^{-1}$. We obtain a maximum count rate of 220~MHz, which is a factor of 2--3 lower than obtained by our previous UNIC-SPD built with discrete components~\cite{Fan2023}.  We attribute the cause of this difference primarily to the 3-ns hold-off time applied in the current module.
The protective series resistor 47~k$\Omega$
used in the module could also play a role in limiting the maximum count rate~\cite{koehler2018best} .  
Nevertheless, the measured count rate is sufficient to allow quantum key distribution beyond 10~Mb/s~\cite{yuan2018}.
As the incident flux exceeds 100 photons per pulse, the detector count rate starts to drop.  In this regime, the avalanche probability within each gate approaches unity. However, the APD's output before the UNIC is close to sinusoidal and the avalanche signals will therefore be rejected by the UNIC circuits along with the capacitive response. 

\begin{figure}[h]
    \centering
\includegraphics[width=0.75\textwidth]{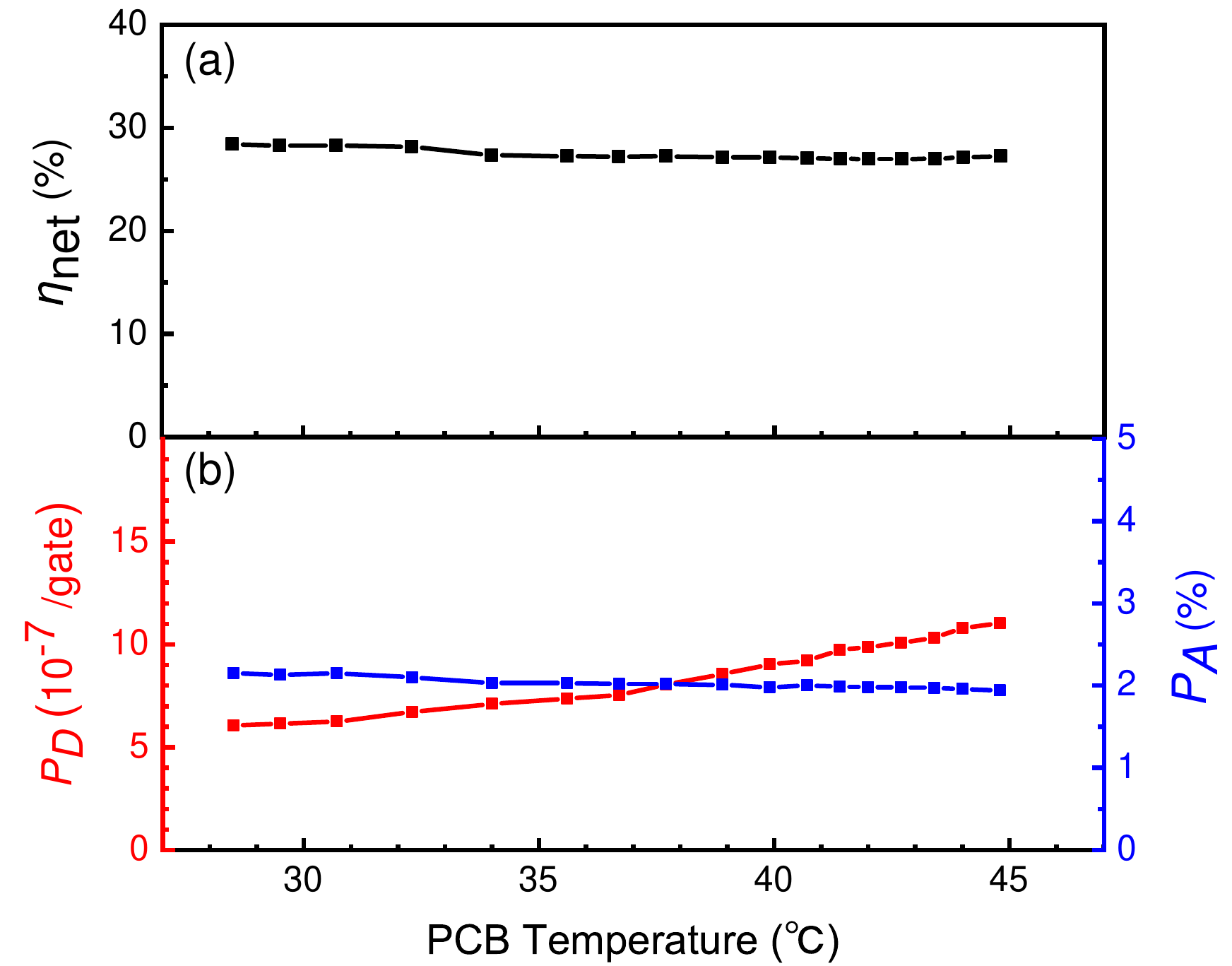}
    \caption{Performance dependence on the PCB temperature. (a) Net photon detection efficiency ($\eta_{net}$)   and (b) dark count probability ($P_D$) and afterpulse probability ($P_A$) versus the PCB temperature measured by the onboard DTS. }
    \label{Fig.5}
\end{figure}

Finally, we test the impact of the PCB temperature on the detection performance. We control the temperature by adjusting the external cooling fan speed to the metal case of the UNIC-SPD module.  During the measurement, the FPGA reads the temperature from the DTS and automatically adjusts the varactor diode and the VCA to compensate for the temperature-dependent group delay of the SAW filter and thus ensure an optimal performance of the readout circuit.  During this test, the bias and temperature settings for the APD are left unchanged, although the actual temperature and the breakdown voltage of the APD could subject to variation by ambient temperature. Over a temperature range of 28 $-$ 45~$^\circ$C,  
we measure the detector performances with results summarised in Fig.~\ref{Fig.5}.  
Over the entire temperature range, the detection efficiency decreases slightly from 28.5~\% to 27.3~\%, corresponding to less than 3~\% relative change.  Simultaneously, the afterpulsing probability maintains a good stability within 1.9 $-$ 2.1~\%.  Relatively, the dark count probability records stronger relative variation, but still maintains a low value of $1.1\times 10^{-6}$ per gate at the highest PCB temperature. These results validate that our UNIC-SPD module can operate stably over a wide range of temperatures. We note in Fig.~\ref{Fig.5}(b) that $P_D$ and $P_A$ exhibit opposite dependencies with PCB temperature, which can be explained by an increase of the actual temperature of the APD device.  
This explanation is consistent with the slight drop in the detection efficiency.
In future work, it is possible to add APD temperature compensation into the FPGA control.

We compare the performance of the UNIC-SPD module with our previous generation~\cite{Fan2023} of UNIC detectors which were implemented with bulk biasing and discrimination equipment. Both detectors incorporate the same model of InGaAs APDs without performance screening. At $\eta_{net} = 30~\%$ and an APD temperature of -15$^\circ$, two detectors have very similar afterpulsing probabilities while the UNIC-SPD module exhibits even 6 times lower dark count probability.  While the observed performance difference is still within variations among InGaAs APD devices, it is fair to conclude that the modularisation and miniaturisation does not cause performance degradation.  

Table~\ref{tab:Table 1} further compares our UNIC-SPD module with two iterations of gigahertz-gated InGaAs/InP SPD modules~\cite{Jiang:17,Jiang2018} operated in 1.25 GHz gating frequency that  integrated a monolithical readout circuit for miniaturisation.
Our module is nearly 4 times smaller in volume than the prior state-of-the-art~\cite{Jiang2018}. Simultaneously, this size reduction does not bring performance deterioration.   Our module reports comparable dark count probability, and a considerably lower afterpulsing probability that is helped by its higher APD temperature.  

\section{Conclusion}

In conclusion, we have developed a fully-integrated InGaAs/InP detector module of a size of just 8.8 × 6 × 2 $\rm {cm}^3$.  We use our previous UNIC techniques for the APD signal readout, but add an automatic temperature compensation to ensure an optimal performance over a wide ambient temperature range. With a 1.25~GHz clock input, the module is characterised to have comparable performance to its counterpaert built with bench-top equipment. 
The UNIC-SPD exhibits excellent performance with a net detection efficiency of 30~\% at an afterpulse probability of 2.4~\% under 3 ns hold-off time. The compact size and state-of-the-art performance allows our UNIC-SPD module a huge potential for single-photon imaging and high speed quantum key distribution.

\subsection*{Author Contributions} 

Y. Fan designed the module and performed the experiment together with Z. Yan and T. Shi. 
Z. Yan and T. Shi prepared the manuscript, and all authors contributed to data analysis and discussions. Z. Yuan supervised the project. 

\subsection*{Funding}
This work was supported by the National Natural Science Foundation of China (Grant number:  62250710162).

\subsection*{Conflicts of Interest}

The authors declare that there are no conflicts of interest related to this article.

\subsection*{Data Availability}
Data underlying the results presented in this paper are not publicly available at this time but may be obtained from the authors upon reasonable request.  

\printbibliography
\end{document}